\pdfoutput=1
\documentclass[fleqn,usenatbib,usecolumn]{mnras}

\usepackage[british]{babel}             
\usepackage{newtxtext}                  
\usepackage[slantedGreek]{newtxmath}    

\usepackage[T1]{fontenc}
\usepackage{ae,aecompl}
\usepackage[T1]{fontenc}                

\usepackage{graphicx}	
\usepackage{amsmath}	
\usepackage{amssymb}	
\usepackage{enumitem}
\usepackage{physics}
\usepackage[dvipsnames]{xcolor}

\newcommand{\Ms}{${M_\ast}$}
\newcommand{\Mstot}{$M_{\ast, \rm tot}$}

\newcommand{\Msol}{$\rm{M_\odot}$}

\newcommand{\HI}{$\rm{HI}$}
\newcommand{\HII}{$\rm{H_{2}}$}
\newcommand{\Ha}{$\rm{H \alpha}$}
\newcommand{\Hb}{$\rm{H \beta}$}
\newcommand{\Sersic}{S\'ersic }
\newcommand{\fgas}{$f_{\rm{gas}}$}
\newcommand{\Mgas}{${M_{\rm gas, fib}}$}
\newcommand{\DMs}{$\rm{\upDelta SF_{MS}}$}
\newcommand{\Msfr}{${M_{\ast, \rm tot}}-{\rm SFR_{tot}}$}
\newcommand{\SFR}{$\rm{SFR_{tot}}$}
\newcommand{\ali}[1]{\textcolor{Black}{#1}}

\newcommand{\rob}[1]{\textcolor{Black}{#1}}

\title[Gas content and quenching in SDSS DR7]{Towards a deeper understanding
of the physics driving galaxy quenching -- inferring trends in the gas content 
via extinction}

\author[J. M. Piotrowska et al.]{
Joanna M. Piotrowska,$^{1,2}$\thanks{E-mail: jmp218@cam.ac.uk}
Asa F. L. Bluck,$^{1,2}$
Roberto Maiolino,$^{1,2}$
\newauthor
Alice Concas$^{1,2}$
and Yingjie Peng $^{3}$
\\
$^{1}$Kavli Institute for Cosmology, University of Cambridge, Madingley Road, Cambridge, CB3 0HA, UK \\
$^{2}$Cavendish Laboratory, Astrophysics Group, University of Cambridge, 9 JJ Thomson Avenue, Cambridge, CB3 0HE, UK \\
$^{3}$Kavli Institute for Astronomy and Astrophysics, Peking University, Yi He Yuan Lu 5, Hai Dian District, Beijing 100871, People's Republic of China \\
}

\date{Accepted XXX. Received YYY; in original form ZZZ}

\pubyear{2019}

\begin{document}
\label{firstpage}
\pagerange{\pageref{firstpage}--\pageref{lastpage}}
\maketitle

\begin{abstract}
In order to investigate the importance of 
different proposed quenching 
mechanisms, we use an~indirect method to estimate gas masses for 
$\sim$62,000 SDSS DR7 galaxies. We infer gas surface
densities from dust column densities as traced by extinction
within the fibre, applying 
a~metallicity correction to account for varying dust-to-gas ratios. 
We find that both gas fraction and star formation efficiency (SFE) 
decrease \rob{moving} away from the star forming main sequence 
(MS) \rob{towards quiescence} for all galaxy masses. We further show 
that both quantities 
correlate similarly strongly with the departure from the MS, 
implying the need for any physical model of quenching to invoke 
a change in \textit{both} gas fraction and SFE. Our results call for a~better 
understanding of the physical processes driving the decrease in star formation 
efficiency, which has received relatively little attention in the theory of 
quenching until now.
\end{abstract}



\section{Introduction}
\label{sec:intro}
Understanding the physical processes responsible for
ceasing star formation is one of the long-standing
questions in the area of galaxy evolution.
Quenching has been shown to correlate well with the total stellar mass of
galaxies (e.g.\
\citealt{Baldry2006, Peng2010}),
their environment 
(e.g.\ \citealt{Bosch2008, Peng2012, Peng2015, Woo2013, 
Bluck2016})
and morphology (e.g.\ \citealt{Cameron2009, 
Bluck2014, Bluck2019}).
A range of phenomena which regulate the gas reservoirs 
of galaxies were suggested as potential causes
of these relations. The processes include gas removal through outflows
(e.g.~\citealt{Maiolino2012, Cicone2014}),
\rob{or galaxy starvation as a consequence of} 
heating of galactic halos by AGN jets \rob{or winds}
(e.g.~\citealt{Croton2006, Fabian2012}) 
or virial shocks
(e.g.~\citealt{Dekel2006, Dekel2009}), 
as well as gas and
halo stripping due to ram pressure exerted on satellites
(e.g.~\citealt{Kenney2004}).

Ultimately, galaxies may quench
either due to a lack of fuel or 
a decrease in star formation efficiency
(i.e. an increase in the depletion time).
To differentiate between these two possibilities
we need to measure
the dense neutral gas within galaxies.
The most direct 
probes of gas content are through
the observations of the \HI\ 21~cm line and the inference
of \HII\ masses from observed CO rotational transition lines 
(see \citealt{CarilliWalter2013} for a review). 

Multiple studies have been conducted to measure the neutral gas
content in galaxies, including recently
\cite{Haynes2018} and \cite{Parkash2018} for \HI\ and 
\cite{Cicone2017}, \cite{Saintonge2017} and \rob{\cite{Tacconi2018}} 
for \HII. However, observing
the faint gas emission typically requires 
long exposure times on premier facilities, thus limiting
the sample sizes to only a~few hundred detections.
Alternatively, less direct
methods utilize the far infrared
dust emission and empirically calibrated
gas-to-dust conversion factors
(e.g.\ \citealt{Santini2014, Scoville2017})
to infer gas masses. 
However, these measurements typically also require 
multiband data in order to remove the degeneracy 
between dust mass and temperature, or else
rely on assumptions 
about the latter.

Given that galaxy quenching
depends on a~range of parameters such as mass, morphology 
and environment, it is difficult to investigate the subtle
interplay of parameters in sample
sizes of only a~few hundred galaxies.
Hence, in order to make progress 
in understanding the physical drivers of quenching we 
must find alternative reliable methods
of estimating gas masses for large galaxy samples,
to complement direct detections.

In this work we utilise the correlation between 
the properties of optical spectra and 
neutral gas content
(e.g.,\ \citealt{Bohlin1978, 
Concas2019}).
In our approach, similar to that of, e.g., 
\cite{Brinchmann2013} and \cite{Barrera2018},
we make use of the attenuation caused
by dust to infer gas column densities in the SDSS 
galaxy spectra.
With this method we trade off accuracy for 
more than a~factor of a~hundred increase in
sample size \rob{(i.e.\, more than 60,000 galaxies)}. 
This provides an unprecedented view of the neutral gas 
content of galaxies in the local Universe.

This letter is structured as follows: in 
Sections~\ref{sec:data}~and~\ref{sec:methods:gas-masses} 
we describe
our sample and gas mass inference method.
Section~\ref{sec:results} describes our results, followed
by discussion in Section~\ref{sec:discussion}. We present
our summary and conclusions in Section~\ref{sec:conclusions}.

\vspace{-0.5cm}
\section{Data and sample selection}
\label{sec:data}

In our analysis we make use of two publicly available derived 
data products for the
SDSS DR7 (\citealt{Abazajian2009}): 
the MPA-JHU\footnote{
\url{https://wwwmpa.mpa-garching.mpg.de/SDSS/DR7/}}
release 
of spectral measurements and the morphological catalogue 
of single \Sersic profile
fits to photometry by \citet{Simard2011}\footnote{\url{ftp://ftp.hia.nrc.gc.ca/pub/users/lsd/astroph/ApJS_SDSS_2011/}}.

After matching entries from both catalogues 
we arrive at a~total number of 
626,062 individual objects, which comprise our parent sample.
Each galaxy has an
associated in-fibre estimate of star formation rate \citep{Brinchmann2004} 
and stellar mass (following the \citealt{Kauffmann2003} approach) along
with aperture corrections to total quantities as prescribed by 
\citet{Salim2007}. All objects have a~measurement of the 
\Ha$(\lambda 6563)$, \Hb$(\lambda 4861)$, 
[\ion{O}{III}]$(\lambda 5007)$,
[\ion{N}{II}]$(\lambda 6584)$ and [\ion{S}{II}]$(\lambda\lambda(6717,6731))$
emission line fluxes, 
provided by the MPA-JHU release as well. They also have an~estimate
of the effective radii and bulge ellipticities from the single 
\Sersic profile fits.

We select our sample by
applying the following criteria:
\begin{itemize}[topsep=0pt, leftmargin=.2cm]
\item a~galactic disk inclination cut of $\lesssim 45^{\circ}$, 
which corresponds 
to an~axial ratio cut of $b/a > 0.7$,
\item a~limit in total stellar mass of \Mstot$>10^8$ \Msol,
\item S/N cuts of $>2$ for \Hb, $>6$ for \Ha\ and $>3$ for 
[\ion{O}{III}] and [\ion{N}{II}] emission line fluxes
\end{itemize}
We also remove potential AGN candidates, defined as lying 
above the \citet{Kewley2001} line in the NII-BPT diagram
\citep{BPT1981}.
Application of our selection cuts results in 61,158 objects in 
our final sample with a median redshift of 0.08 and
MAD of 0.04.

The inclination cut removes edge-on galaxies,
rejecting extremely dust obscured objects 
due to \rob{geometry effects rather than pure dust content}. 
We require an S/N cut on \Hb\ and \Ha\
to infer $E(B-V)$ and our choice of low \Hb\ 
cut aims at striking
a balance between accuracy and completeness.
Remaining S/N cuts are imposed by our choice of metallicity
calibrator.
\rob{Finally, we remove the AGN sources because they are not 
likely to sample the dust in the host galaxy,
as shown by e.g.\, \cite{Concas2019}, 
since the nebular lines
are primarily emitted in the ionisation cones of the narrow-line
region rather than in the galactic disk.}
By imposing these selection criteria we preferentially
discard the passive sequence, 
reducing sample
completeness to below 10\% in this region of the \Msfr\ space. 
Hence, we cannot draw
robust
conclusions about the passive sequence and restrict our 
primary discussion
to objects ranging from the green valley to starbusts.

\vspace{-0.5cm}
\section{Estimating \rob{gas} masses} 
\label{sec:methods:gas-masses}

\begin{figure}
	\includegraphics[width=\columnwidth]{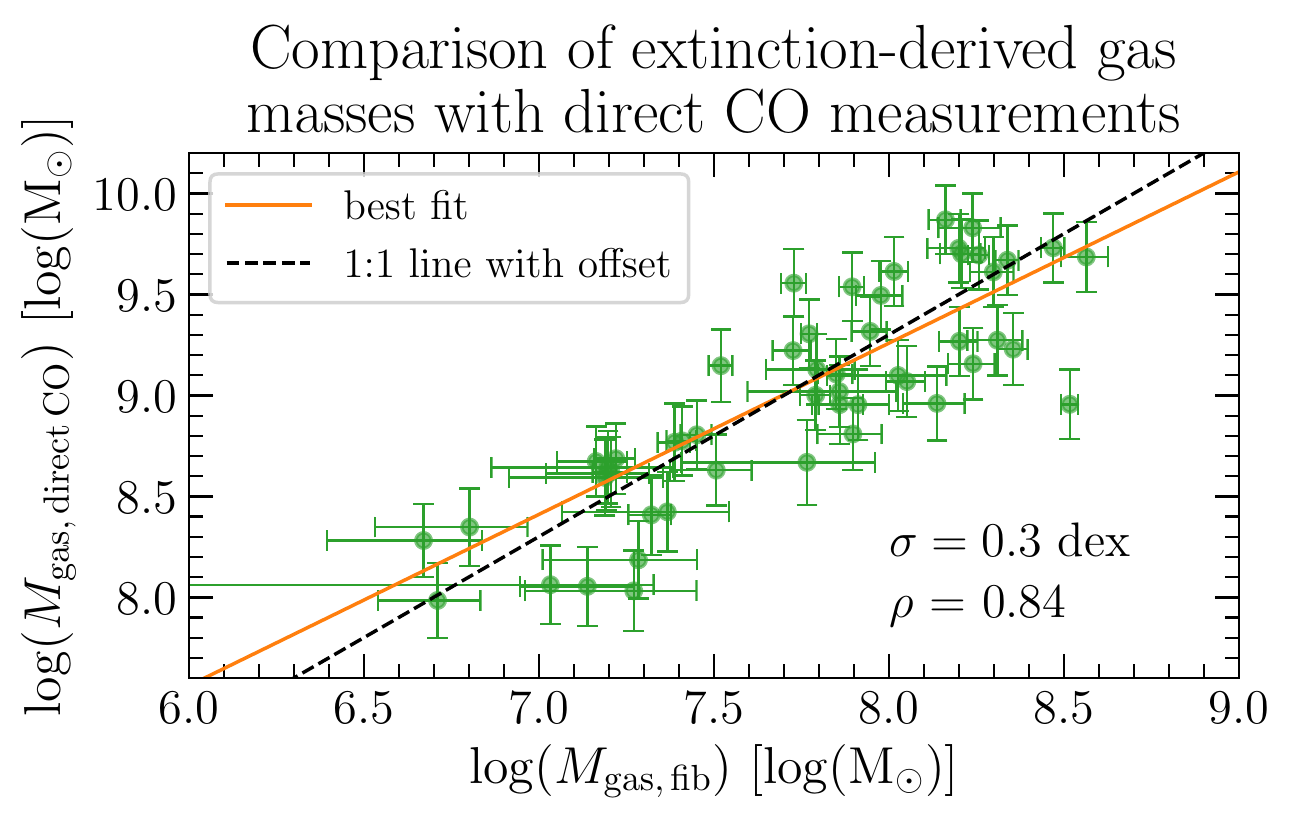}
	\vspace{-0.5cm}
	\caption{Comparison between gas masses obtained from
	direct CO observations by \citet{Saintonge2017} and this work
	for 57 matched objects. Horizontal errorbars are result of 
	propagation of uncertainties in fluxes and eq.~(\ref{eq:sigmaH}).
    The values exhibit a~linear relation	
	with a~best fit plotted as the orange solid line. 
	A~unit slope best
	fit is shown as black dashed line with an intercept value of 1.30.	
	This strong correlation ($\rho=0.84$)
	and moderate scatter ($\sigma=0.3$ dex)
	support the use of dust reddening 
	as proxy for estimating
	the gas content of the SDSS galaxies.}
    \label{fig:1}
\end{figure}

We utilise dust reddening
to infer the average column densities of neutral
hydrogen from the SDSS optical spectra. 
More specifically, we use \ali{an updated version
of an~empirical relation between the neutral hydrogen 
number density and colour excess $E(B-V)$} 
\citep{Bohlin1978}, 
obtained by \citet{Gudennavar2012}:
\begin{equation}
N(\rm{H}) = (6.12 \pm 0.20)\ E(B-V) 
+ 0.19\  \ 10^{21}\rm{atom\ cm^{-2}\ mag^{-1}} ,
   \label{eq:bohlin}
\end{equation}
where $E(B-V)=A_V/R_V$ and $R_V \sim 3.1$ 
for the Milky Way. \rob{This relation was derived in the Galaxy
for lines of sight dominated by atomic hydrogen. In our method 
we assume that the relation extends to 
total column densities which include regions of dense, molecular
phase.} We explicitly test this assumption later in this section.

Since equation~(\ref{eq:bohlin}) was
derived from observations within the Milky Way, we need to 
account for the dependence of the dust-to-gas ratio on
gas-phase metallicity (e.g.\, \citealt{Issa1990, RemyRuyer2014}).
For this reason we include a~metallicity correction 
in our estimation of gas masses, assuming a~linear dependence 
of the dust-to-gas ratio on the metal content in the gas.
In order to implement this correction we estimate metallicity across the 
entire \Msfr\  plane using the O3N2 calibration 
of nebular emission line ratios extended to non purely star-forming 
regions by \citet{Kumari2019}. 

Finally, we make a~choice of dust geometry for all objects.
\rob{We assume an~effective foreground screen 
containing half the total amount of dust in a~given galaxy
(e.g.,\, \citealt{Imara2007}, 
\citealt{Barrera2018}).
A different choice of geometry would influence the conversion
factor between the gas mass surface density $\Sigma_{\rm gas}$
and $A_V$, affecting the quantitative result but not the
qualitative trends with other parameters.}

Thus, the final relation we use to estimate $\Sigma_{\rm gas}$,
assuming a~uniform foreground dust screen 
and linear metallicity
dependence in the gas-to-dust ratio, has the form:
\begin{equation}
\Sigma_{\rm gas} = \qty[(31.6 \pm 1.0)A_V + 1.0]
\qty(\flatfrac{Z_\odot}{Z})\ 
{\rm M_\odot {pc^{-2}}}\, .
\label{eq:sigmaH}	
\end{equation}

In order to estimate $A_V$ we
compare the observed ${F_{\rm H\alpha}/F_{\rm H\beta}}$ 
to its intrinsic
\rob{Case B} value of 2.86 (\citealt{Hummer1987})
via  the following formula:
\vspace{-6pt}
\begin{equation}
A_\lambda = \frac{K_{\lambda}}{K_{\rm H\beta} - K_{\rm H\alpha}} 
2.5 \log_{10}\Bigg(\frac{{F_{\rm H\alpha}}/{F_{\rm H\beta}}}{2.86}\Bigg)\, ,
\label{eq:AV}
\end{equation}
where $K_{\lambda}=A_{\lambda}/A_V$ is taken from any 
attenuation law of 
choice. In our study, we choose the parametrisation
provided by \citet{Cardelli1989} for a~Milky Way
\rob{attenuation} curve of ${R_V \sim 3.1}$, 
where $K_V=1$ by construction, ${K_{H\alpha}=0.817}$
and ${K_{H\beta}=1.164}$. We check that the choice of
of the attenuation curve does not impact our results by using 
the \cite{Calzetti2000} curve as an additional test. 
Although we find 
small quantitative differences in the resultant gas masses,
all trends and hence conclusions are identical.
Finally, to obtain a~gas mass 
estimate, we integrate 
equation~(\ref{eq:sigmaH}) over the physical area covered by
each 3" diameter fibre.

In order to validate our indirect method we select objects
in our final sample with xCOLDGASS detections 
\citep{Saintonge2017}
to compare gas masses inferred from reddening with those
obtained through CO-to-\HII\ conversion. Fig.~\ref{fig:1}
presents a~comparison between the 57 objects present
in both samples, which
show a~strong linear relation with Spearman's rank correlation
coefficient of $\rho=0.84$. 
The in-fibre gas mass estimates are offset to lower values,
which is expected because the integrated
CO measurements probe the total gas mass in a~given
galaxy, whereas our approach estimates the in-fibre mass.
The best linear fit \rob{in the log-log plane} 
(orange solid line) 
yields a~slope of $0.85\pm0.08$~dex and an intercept
of $2.5\pm0.6$~dex, while if we fix the slope to unity we obtain an
offset of $1.30\pm0.04$~dex with an rms scatter of $0.3$~dex around
the one-to-one line. This test is encouraging and supports
the use of our gas mass estimates to investigate trends in 
the SDSS DR7 sample. 

In order to check whether our main results depend on
the choice of calibration we also explored a prescription
without the metallicity correction and a preliminary empirical
calibration between ${F_{\rm H_\alpha} / F_{\rm H_\beta}}$
and $\Sigma_{\rm gas}$ from the
ALMA-MaNGA QUEnching and STar formation survey,
(L. Lin et al. in prep.) by Concas et al. 
(2019, in prep.). We find that all of our main conclusions are preserved,
despite some differences in the inferred gas masses of individual 
galaxies.

\vspace{-0.5cm}
\section{Results}
\label{sec:results}

\begin{figure}
	\includegraphics[width=0.93\columnwidth]{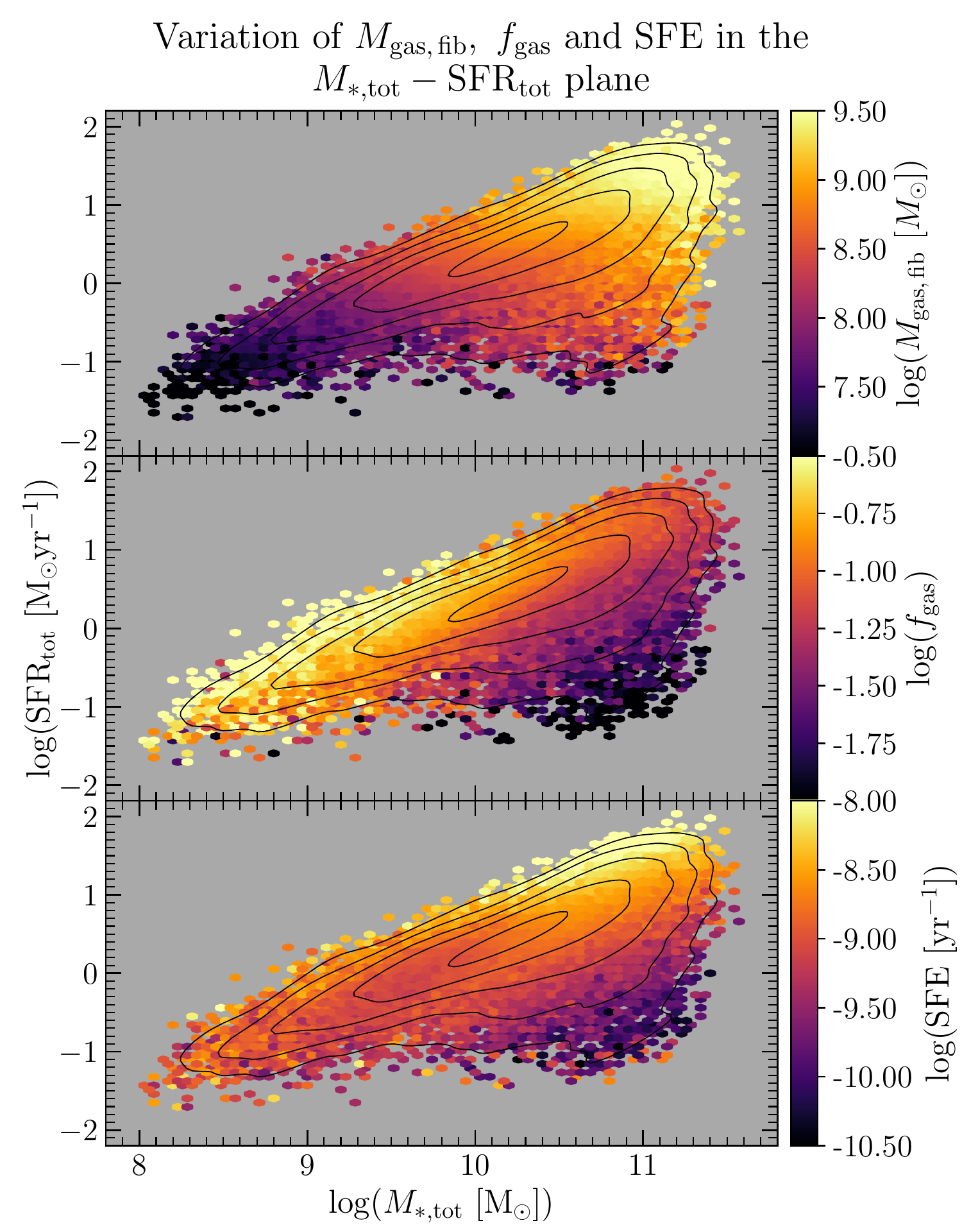}
	\vspace{-0.3cm}
	\caption{Top: in-fibre gas mass (\Mgas), 
    middle: in-fibre gas fraction (\fgas) and bottom: 
    in-fibre  star formation efficiency (SFE) variation across
	the \Msfr\ plane, presented in hexagonal bins colour-coded
	by the median value in each bin. Black contours 
	indicate the number
	density of galaxies in the plane and are evenly spaced in
	logarithmic space. 
	}
    \label{fig:2}
\end{figure}

In order to analyse trends with global galactic properties,
we compute the gas fraction ${f_{\rm gas}=M_{\rm gas,\, fib}
/ M_{\rm \ast,\, fib}}$
and star formation efficiency 
${\rm SFE} = {\rm SFR_{fib}}/M_{\rm gas,\, fib} = 1/\tau_{\rm dep}$,
where the subscript `fib' denotes in-fibre quantities. The former
informs us about the fuel available for star formation
while the latter how efficient the gas is at collapsing
to form new stars. We focus
on the in-fibre quantities, as our gas mass inference
is limited to the spectroscopic measurements 
(i.e.\, the central 3" of each galaxy). 
The use of total
quantities would require strong assumptions to compute
aperture corrections.

In Fig.~\ref{fig:2} we present the variation in \Mgas\ in the top panel;
\fgas\ in the middle panel and SFE in the bottom panel across
the \Msfr\ plane. The results are shown in hexagonal bins
colour-coded by the median value of each respective quantity.
Black contours trace the density
of objects in the plane and aid in visualising the distribution
of the sample, dominated by the high-SFR objects. 
Across the Main Sequence (MS), 
\Mgas\ shows a~smooth increase of $\sim3$~dex 
with increasing total stellar mass, accompanied by a mild
decrease in gas fraction and an increase in star formation
efficiency.
Moving off the MS towards lower SFRs, \Mgas\ slightly 
decreases with decreasing \SFR\ at constant \Mstot, while
both \fgas\ and SFE show a dramatic drop in values
of $\sim1$~dex and $\sim1.5$~dex, respectively.
All quantities also exhibit an increased scatter towards the
passive sequence due to the deteriorating statistics.
 
At a first qualitative glance both \fgas\ and SFE show
a~dramatic variation with increasing deviation from the MS.
In order to quantify these trends, we further
split our sample into four bins of \Mstot\
and then for each mass bin examine the variation 
in \fgas\ and SFE as a~function of deviation 
from the MS. 
We define the deviation from the MS (\DMs) as 
distance from the star forming Main Sequence parametrised
by a least squares fit to the galaxy
sample with ${{\rm sSFR}=\flatfrac{\rm SFR}{M_{\ast, \rm tot}} > 10^{-10.8}
{\rm yr}^{-1}}$ (which excludes the green valley and passive 
sequence). Our fit broadly agrees with the definitions by 
{\cite{Renzini2015} and \cite{Bluck2016}.}

\begin{figure}
    \begin{center}
\includegraphics[width=0.93\columnwidth]{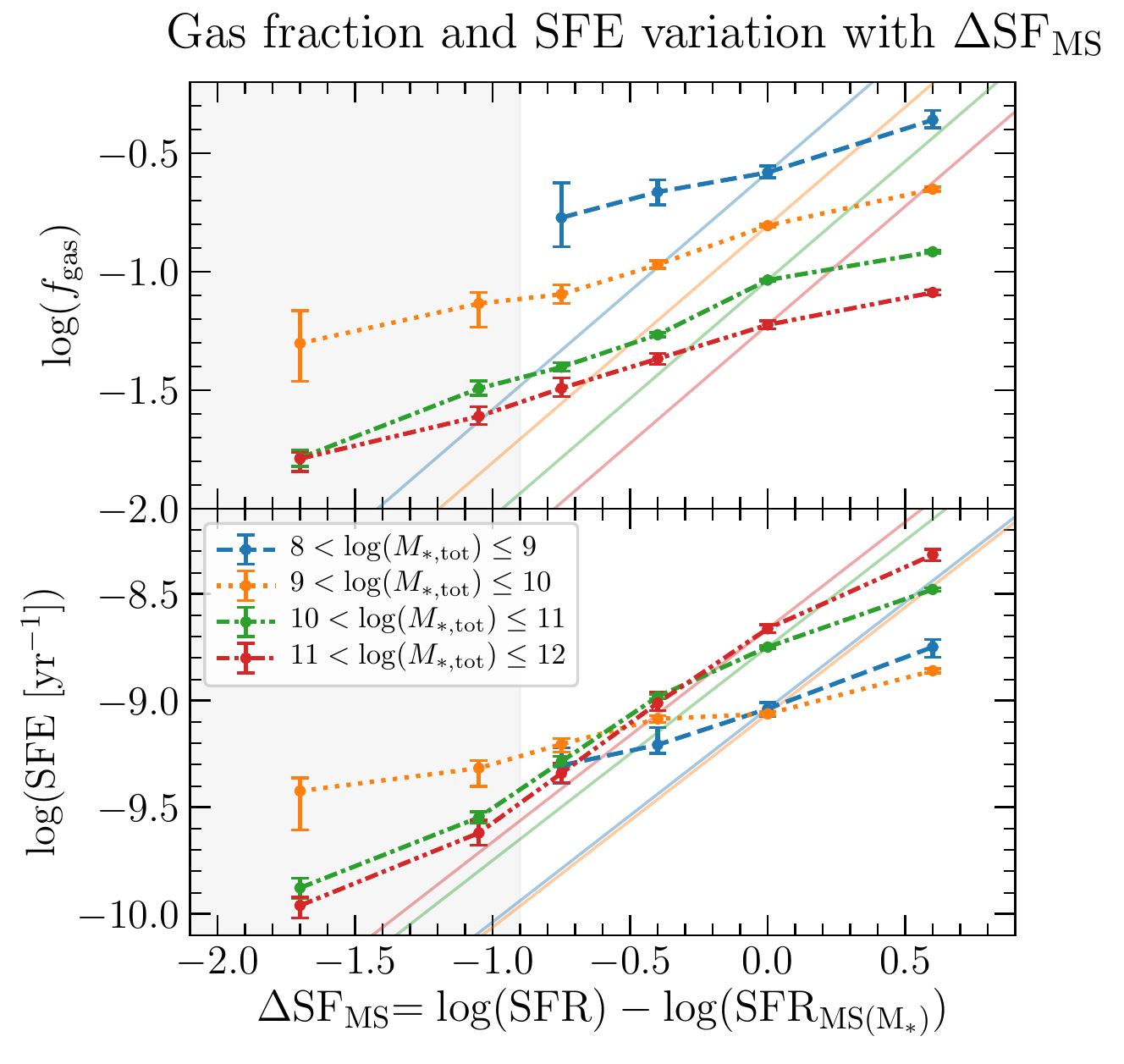}
    \end{center}
\vspace{-0.5cm}
\caption{Top panel: gas fraction (\fgas)
and bottom panel: star formation efficiency (SFE) 
plotted as functions 
of distance from the MS in four bins of total stellar 
mass. Each point corresponds to a~median value with 
a~bootstrapped error estimate.
The grey shaded area covers the range in \DMs\ with $\lesssim10\%$
completeness in each bin.
Both quantities show a~clear decrease in value with increasing
departure from the MS. The solid lines show the variation in either
\fgas\ (top panel) or SFE (bottom panel)
required to produce the observed variation in 
\DMs\ whilst keeping
the other parameter constant.
The fact that neither $f_{\rm gas}$ nor SFE
trace these lines indicates that both parameters correlate
with quenching.
}
\label{fig:3}
\end{figure}

Fig.~\ref{fig:3} presents \fgas\ (top panel) and SFE (bottom panel)
varying as functions of \DMs, where y-values
correspond to medians of the given quantity in each \DMs\
bin, while x-values indicate the bin centres. Associated
95\% confidence bounds on the median are
estimated by bootstrapping and presented as vertical 
error bars. The grey shaded region indicates the range of \DMs\ values
where the completeness in each bin is lower than 10\%.

Both
quantities in Fig~\ref{fig:3} visibly
decrease moving away from the MS 
towards the quiescent region, and increase moving 
away in the opposite direction (towards 
starburst galaxies), for all ranges in \Mstot.
\fgas\ seems to follow similar gradients for
all mass bins, while SFE indicates a~steeper
decrease with decreasing \DMs\ in more massive
systems than in the less massive ones. 
The solid coloured lines in Fig.~\ref{fig:3}
illustrate the variation in \fgas\ and SFE
required to achieve the variation in \DMs\ without
variation in the other quantity.That is, we use
the identity: $\rm{sSFR}=f_{\rm gas} \times {\rm SFE}$, 
fixing SFE to its MS value 
in the top panel, and \fgas\ to its MS value in the 
bottom panel to obtain the expected variation in the
other parameter as a function of \DMs.

If \DMs\ is driven solely by varying quantity A
the trends in quantity A would line up with the
solid coloured lines and the trends in quantity B 
would be entirely flat.
Fig.~\ref{fig:3} shows, however,
that neither \fgas\ nor SFE is in a~perfect agreement 
with the solid lines, suggesting that both parameters are 
significantly correlated with deviation from the MS.
The Spearman correlation coefficient between \fgas\ and
\DMs\ is 0.44, while between SFE and \DMs\ is 0.49.
This result indicates
that as we move down towards the passive sequence,
\textit{it is not only the gas reservoir of a~galaxy which
decreases but also the efficiency with which the gas 
is turned into stars}. 

\begin{figure}
\includegraphics[width=\columnwidth]{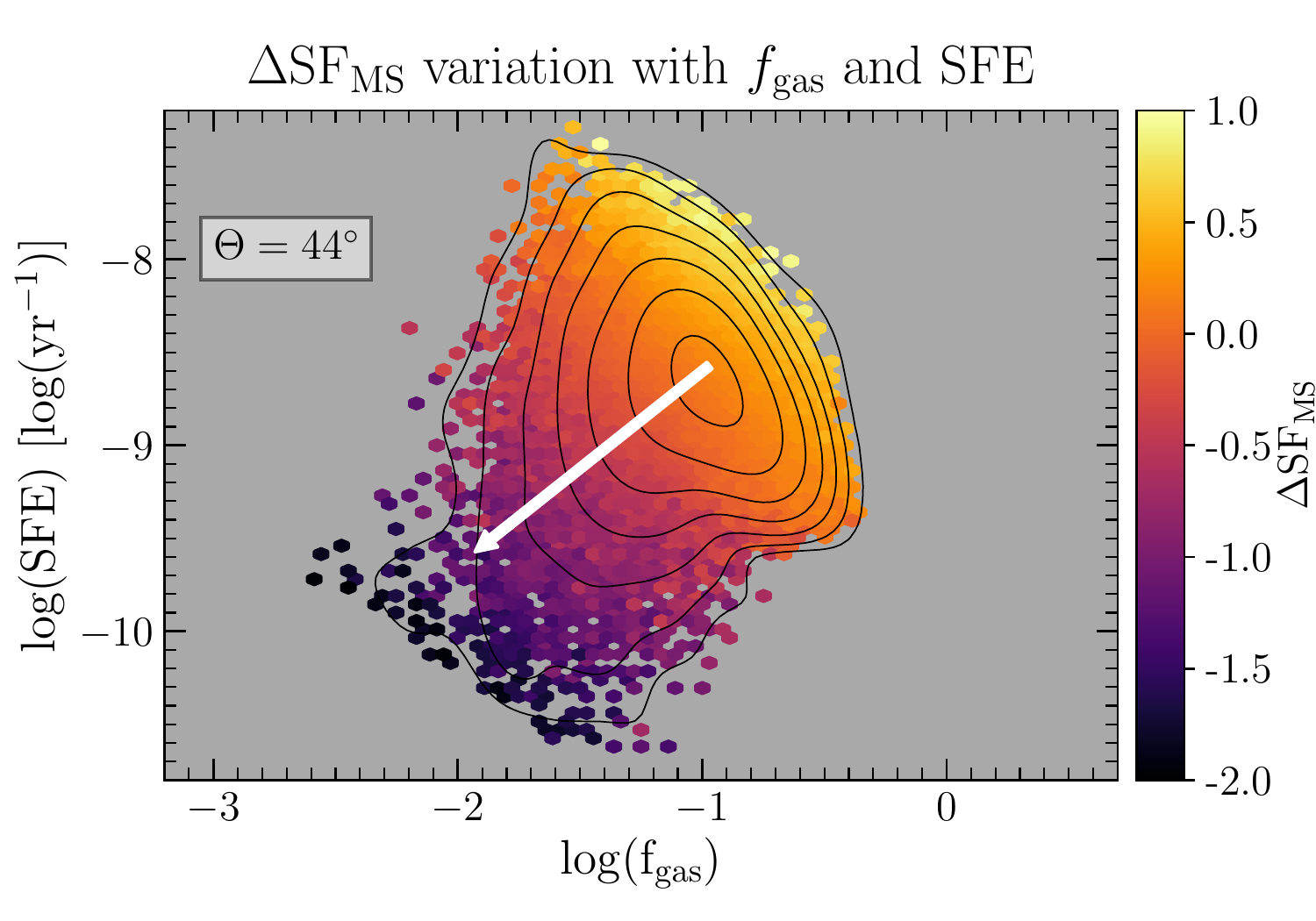}
\vspace{-0.6cm}
\caption{Variation of \DMs\ in the \fgas -SFE plane for galaxies with
	${M_\ast > 10^{10} \rm{M_\odot}}$.
	The colour of each hexagonal bin corresponds to the median value
	of \DMs.
	Black contours indicate
	number density of objects in the plane.
	\DMs\
	becomes increasingly negative
	as we decrease both \fgas\ and SFE. 
	The arrow points in the `quenching' direction,
	i,.e the steepest gradient in the \DMs\ value and is inclined
    at $44^{\circ}$ to the vertical axis.}
\label{fig:4}
\end{figure}

In Fig.~\ref{fig:4} we present the variation in 
\DMs\ in the $f_{\rm gas}-{\rm SFE}$
plane for galaxies with $M_\ast > 10^{10} {\rm M_\odot}$.
We can see from the individual galaxy distribution that there is 
a significant variation in both \fgas\ and SFE in the massive
part of our sample. The plot illustrates 
two quenching avenues, showing that in order to move away 
from the MS at high stellar mass, we need to
move towards both lower SFE and lower \fgas\ values, 
deviating fastest
when we decrease both quantities by an approximately 
equal amount.  
The white arrow points in the `quenching direction'
and its inclination angle is calculated following 
\cite{Bluck2019s}, where partial
correlation coefficients (PCC) between \fgas\ and \DMs\
and SFE and \DMs\ are treated as vector components
of the arrow. 
The PCC between A~and B calculated
when controlling for \mbox{C is given by $\rho_{\rm{AB|C}}$}:
\begin{equation}
\rho_{\rm{AB|C}}=\frac{\rho_{\rm{AB}}-\rho_{\rm{AC}}\cdot\rho_{\rm{BC}}}
{\sqrt{1-{\rho^2_{\rm{AC}}}}\sqrt{1-{\rho^2_{\rm{BC}}}}},
\end{equation}
where $\rho_{\rm{AB}}$ is a~Spearman's rank correlation coefficient
between quantities A~and B. The inclination of the arrow is
then given by $\Theta$:
\vspace{-6pt}
\begin{equation}
\Theta=\tan^{-1}\qty(\frac{\rho_{\rm{CB|A}}}{\rho_{\rm{AB|C}}}),
\label{eq:theta}
\end{equation}
where $\Theta$ is measured from the $Y$-axis. 
In Fig.~\ref{fig:4}
$\Theta \sim45^{\circ}$, as expected from the relationship
between \fgas, SFE and \DMs. Thus, quenching depend roughly
equally on a reduction of gas and an increase in depletion time.

\vspace{-0.5cm}
\section{Discussion}
\label{sec:discussion}

The main conclusion of our analysis
is that both \fgas\ and SFE decrease with deviation from the Main
Sequence, in our selected sample of emission line 
galaxies. Similar trends to those shown in Fig.~\ref{fig:3} are also seen in smaller samples of more direct
measurements, including in xCOLDGASS 
(\citealt{Saintonge2017}),
PHIBSS (\citealt{Tacconi2018}), and ALMA observations of the long
wavelength dust continuum (\citealt{Scoville2017}).
More recently, \cite{Ellison2019s}
also draw similar conclusions about the importance
of SFE in quenching on resolved scales from a sample
of~34~galaxies~observed~with~ALMA.

It is important
to recognise the caveats associated with our results.
First of all, our analysis is limited to the central 1.5" radii
(on average $0.65\pm0.25$
of the effective radii) of galaxies
sampled by the SDSS fibres. To test whether 
this impacts our results, we split our sample into 
narrow redshift bins and repeat the analysis. Although 
minor quantitative differences are present, all
qualitative trends are preserved in each redshift bin.
We also check for a~tight correlation between the 
global and in-fibre
\Ms\ and SFR (see Fig.~B1 in the online supplementary material)
and hence conclude that the fibre aperture is not
a source of significant bias in our analysis.
We also confirm that our conclusions are identical with respect to the
in-fibre MS, as well as the global MS (studied throughout this letter).
Additionally, 
although large,
our sample still remains incomplete in the passive
sequence, since we require emission line detections for our analysis. As a result,  
we are biased towards the most gas rich
of the passive systems. Consequently, our conclusions are robust throughout the MS and green valley, but are tentative in the passive sequence.
Finally, as described in \cite{Brinchmann2013},
the assumption of a single attenuation curve for the
whole sample is a significant simplification, as this 
may vary away from the MS (e.g.\, \citealt{Salim2019}). 
However, we repeat our analysis with the xCOLDGASS sample
(see Fig.~A1 in the online supplementary material)
and find strong agreement with the trends in Fig.~\ref{fig:3}.
Moreover, these trends are consistent with those found
by \citealt{Tacconi2018}, and thus we are
encouraged to conclude that this assumption does
not strongly impact our main result.

There is a~range of plausible physical mechanisms 
for decreasing the gas reservoirs available for star formation.
One possible channel is the expulsion of gas out of the galaxy
in outflows (e.g.\ \citealt{Maiolino2012,
Cicone2014}), which may be driven by AGN or supernovae.
Additionally, powerful AGN jets and/ or winds may deposit
large quantities of energy in the gas halo 
(e.g.\ \citealt{Croton2006, Fabian2012}),
preventing accretion of cold gas through heating. Virial shocks 
experienced by gas rapidly falling through massive
haloes may also be a significant source of heating (e.g.\ \citealt{Dekel2006, Dekel2009}).
In these scenarios galaxies quench as a~consequence 
of `starvation', for which there is indirect evidence based on a 
metallicity analysis of galaxies (e.g.\, \citealt{Peng2015,
Trussler2018}). Satellite galaxies are subject to additional environmental effects, 
including the removal of their dense gas reservoirs  
by ram pressure stripping (e.g.\ \citealt{Kenney2004})  and `strangulation'
due to stripping of their hot gas halos
(e.g.\ \citealt{Peng2015}).
The importance of environment to quenching 
is supported by correlations
between quiescence of satellites and 
local galaxy density (e.g.\ \citealt{Peng2012}),
satellite location within the group, and the virial halo mass
(e.g.\ \citealt{Woo2013}).

Varying SFE as a~cause of quenching has received
less attention in the literature than varying \fgas.
Interstellar turbulence and, to a~lesser degree,
magnetic fields are proposed as mechanisms 
regulating star formation (e.g.\ \citealt{Federrath2012}),
in which case local SFE would
depend on the Mach number and the virial parameter
of each molecular cloud. 
Cyclic evolution of the interstellar medium
(ISM) between the non-star forming and
star forming states could lead to long
global gas depletion times in 
low gas density environments \citep{Semenov2017},
hence potentially linking the decrease in SFE to the decrease
in \fgas\ (discussed above). 
Alternatively, the decreased SFE values we observe
could be driven by the presence of a~prominent 
bulge, which increases the tidal torques and velocity
dispersion within a~galaxy,
preventing gravitational collapse of the 
giant molecular clouds. This
hypothesis would then link the variation in SFE to 
galaxy morphology, as suggested by \citet{Martig2009}.
Finally, the ISM could also be directly heated
by energy injection from different forms of AGN 
activity and stellar winds at different
stages of evolution, preventing gravitational
collapse and hence decreasing the efficiency of
star formation, without significantly reducing \fgas.

In future work we will attempt to break the degeneracies between the plausible models for quenching outlined above, by utilising the statistical power of our large sample of galaxies with indirect gas mass estimates. More specifically, we will investigate the subtle interplay between mass, morphology, environment and AGN activity in driving variation in \fgas\ and SFE.

\vspace{-0.6cm}
\section{Conclusions}
\label{sec:conclusions}

We infer the gas content of $\sim$62 000 local galaxies
using the reddening
of observed optical spectra in the SDSS DR7. Our choice
of indirect method allows for a~factor of one~hundred increase
in sample size compared to more direct methods, 
trading off accuracy for excellent statistics. 
When validated against CO observations
from \citet{Saintonge2017}, our results show a very strong 
correlation ($\rho=0.84$) and moderate scatter 
($\sigma=0.3$ dex)
between \Mgas\ and $M_{\rm H2}$, encouraging the use
of this method for the entire emission line sample of the SDSS.

In this letter we demonstrate that transition towards quiescence
results in a~significant decrease in {\it both} gas fraction and star formation
efficiency for all galaxy masses, consistent with several previous analyses of 
much smaller samples utilising more direct gas measurements 
(e.g.\, \citealt{Scoville2017, Saintonge2017, Tacconi2018, Ellison2019s}). 
We further show that deviation from the star forming
Main Sequence is similarly strongly correlated with reduction in gas content and lengthening of depletion time. Thus, successful models of quenching must explain both a reduction in gas content and star forming efficiency within transitioning galaxies. 

\vspace{-0.6cm}
\section*{Acknowledgements}

We thank the anonymous reviewer for their
constructive comments which greatly improved this letter.
JMP and YP gratefully acknowledge funding from the MERAC Foundation.
AFLB, RM and AC gratefully acknowledge ERC Advanced Grant 695671
`QUENCH', and support from the Science and Technology Facilities
Council (STFC).
YP acknowledges the National Key R\&D Program of China, Grant 2016YFA0400702 and NSFC Grant No. 11773001.

In order to ensure reproducibility, all analysis is publicly available
at \url{https://hub.docker.com/u/jpiotrowska}.




\bibliographystyle{mnras}
\bibliography{letter} 



\appendix
\section{Comparison to xCOLDGASS}
\label{sec:comparison}

\begin{figure}
\includegraphics[width=\columnwidth]{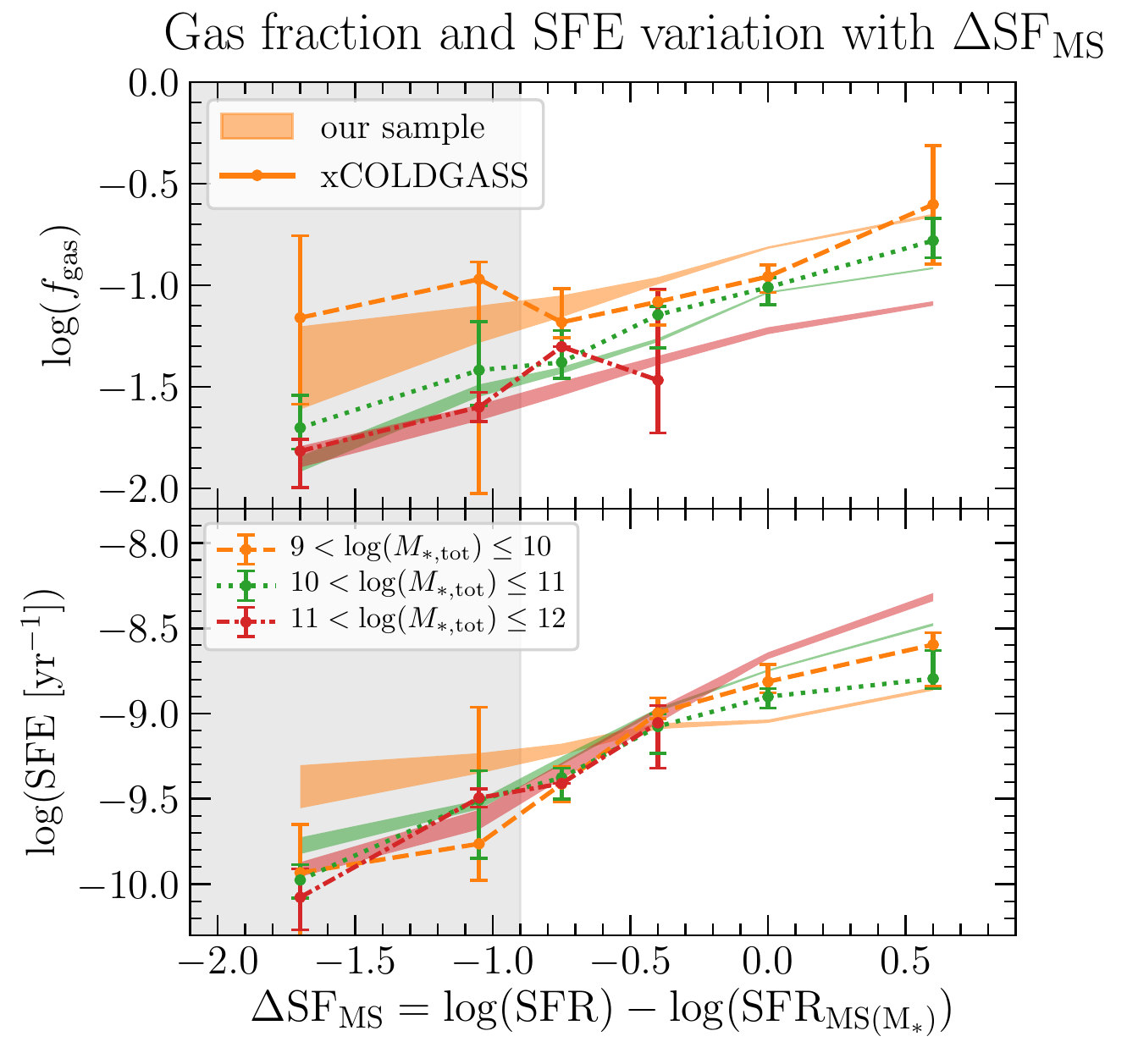}
\vspace{-0.6cm}
\caption{Top panel: gas fraction (\fgas)
and bottom panel: star formation efficiency (SFE) 
plotted as functions 
of distance from the MS in bins of total stellar 
mass. The shaded curves show our data from Fig.~3 in
the letter
while the solid lines with errorbars are computed using
the publicly available xCOLDGASS sample.
The two datasets are in a very good agreement with each other,
further supporting the conclusions drawn from gas mass inference
from optical extinction.}
\label{fig:A}
\end{figure}

\begin{figure}
\includegraphics[width=\columnwidth]{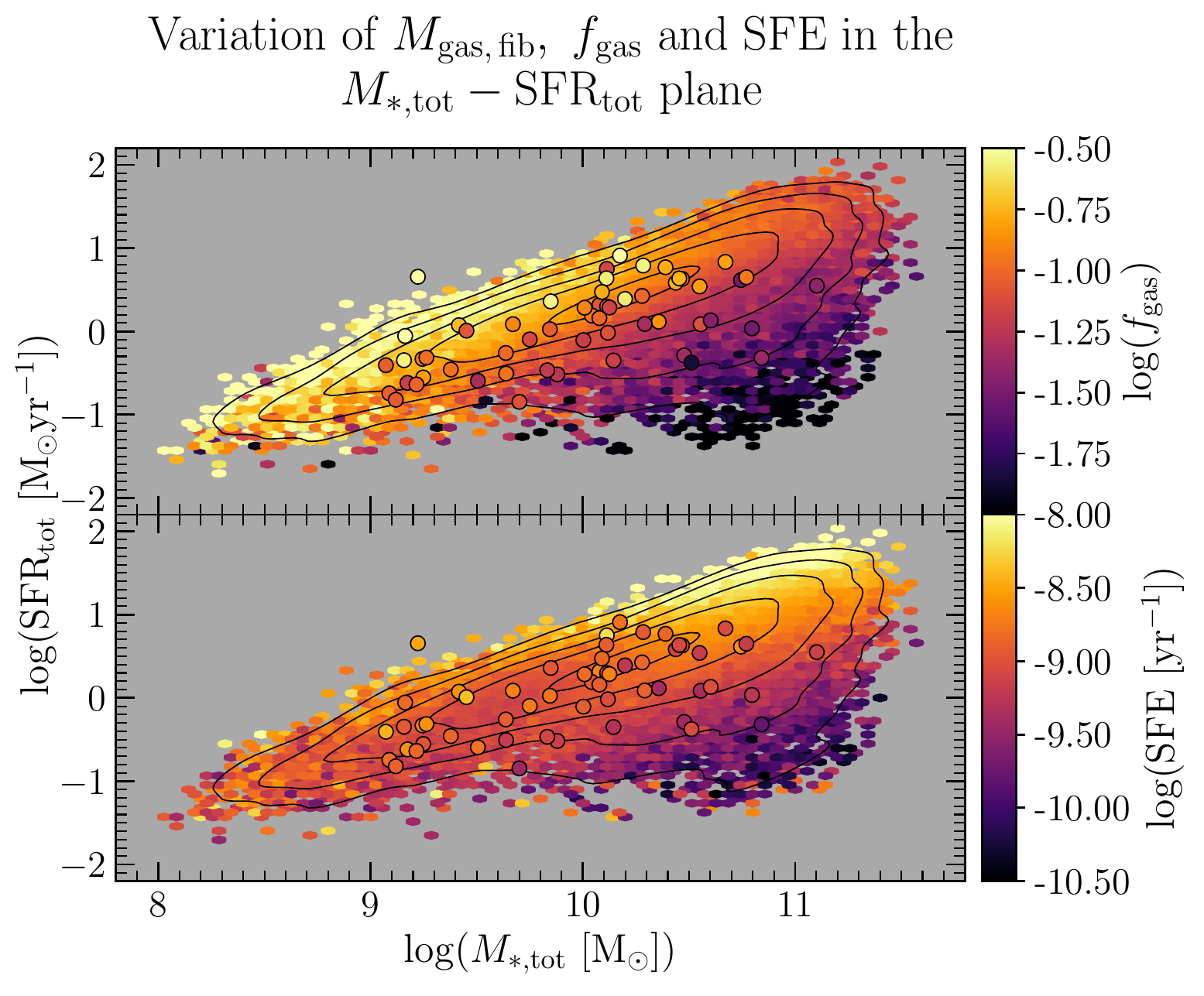}
\vspace{-0.6cm}
\caption{Top: gas fraction \fgas, bottom: star 
formation efficiency (SFE) variation across the
$M_{\ast, \rm tot}-{\rm SFR_{tot}}$ plane. Hexagonal bins
present the same data as Fig.~2 in the letter, while
coloured circles correspond to the matches between
our sample and the xCOLDGASS presented earlier in
Fig.~1. Both panels show a good qualitative
and quantitative agreement between the fibre-inferred
quantities and global measurements based on CO emission.}
\label{fig:B}
\end{figure}

Fig.~\ref{fig:A} presents a direct comparison of trends 
between \fgas, SFE and \DMs\ in the xCOLDGASS and our
selected SDSS sample. The SFR and stellar masses used
to calculate SFE and \fgas\ in both datasets are taken
from the MPA-JHU release for consistency and data is 
binned in the exact same fashion with errors estimated
by bootstrapping in each bin. The colours correspond
to the different mass bins labelled in Fig.~3 in the letter.
Both trends are in a very good agreement with each other.
The direct CO measurements show a decrease in
\fgas\ and SFE as a function of deviation from the MS,
supporting our result obtained with a more indirect
inference from optical extinction. 

The trends in Fig.~\ref{fig:A} are also visible in 
Fig.~\ref{fig:B}, which presents the two lower panels
of Fig.~2 in the letter with 57 matches between our sample
and the xCOLDGASS overplotted as coloured circles. Both
datasets share the same colourmap for a convenient
comparison. The in-fibre inferred quantities (hexagonal bins)
are in good agreement with the total ones (circles) within
the rms scatter of 0.4 dex presented in Fig.~1 in the letter,
showing that the in-fibre analysis does not strongly affect
our conclusions. 

\section{Relation between the in-fibre and global quantities}
\label{sec:comparison}

\begin{figure}
\begin{center}
\includegraphics[width=0.8\columnwidth]{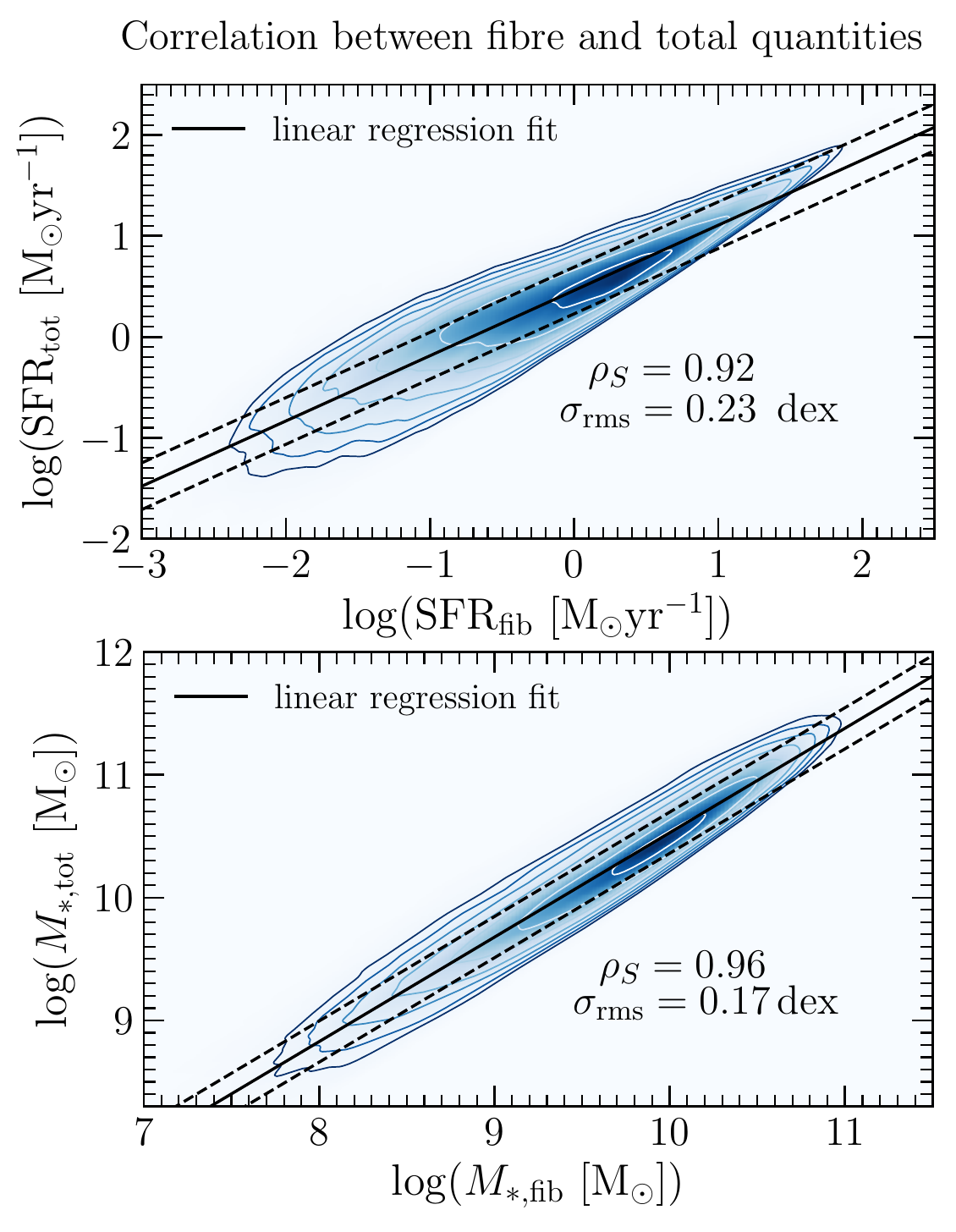}
\end{center}
\vspace{-0.6cm}
\caption{Top panel: SFR, bottom panel: \Ms\ comparison
between the in-fibre and total quantities in our sample.
The in-fibre estimates for both stellar mass and SFR are
very well correlated with the global values with Spearman's
rank correlation coefficients of 0.96 and 0.92 respectively.
The relations are also very tight with the corresponding
scatter around the best linear fit of 0.23 dex for SFR
and 0.17 dex for \Ms. }
\label{fig:C}
\end{figure}

Fig.~\ref{fig:C} further addresses the concern of the influence
of a fixed aperture size on the inferred trends.
It shows strong correlations between the
in-fibre and total quantities within our sample. The
Spearman's rank correlation coefficients of 0.92 and 0.96
as well as small scatter around the linear best fits of
0.23 dex and 0.17 dex for SFR and \Ms\ respectively 
show that the in-fibre analysis is a good proxy for the
global behaviour of quantities of interest in our study.
The trends in Fig.~\ref{fig:C} combined with Fig.~1
in the letter suggest that inferred trends in in-fibre
\fgas\ and SFE should reflect the global ones, as indeed
shown in Fig.~\ref{fig:A}.

\bsp	
\label{lastpage}
\end{document}